\documentstyle[aps,pra,prabib,amsfonts,epsf]{revtex}

\tighten
\twocolumn
\narrowtext

\newcommand{\E}{{\rm e}}
\newcommand{\lkl}{\left(}
\newcommand{\rkl}{\right)}
\newcommand{\vecr}{{\bf r}}
\newcommand{\vecp}{{\bf p}}
\newcommand{\sumCq}{\sum_{C_1,\ldots,C_N; \atop \sum_{q=1}^N q C_q = N}}
\newcommand{\sumCr}{\sum_{C_1,\ldots,C_N; \atop \sum_{r=1}^N r C_r = N}}

\newcommand{\rhos}{\frac{\rho_{\rm{s}}}{\rho}}
\newcommand{\rhost}{\rho_{\rm{s}}/\rho}
\newcommand{\rhon}{\frac{\rho_{\rm{n}}}{\rho}}
\newcommand{\iqm}{I_{\rm{qm}}}
\newcommand{\iclass}{I_{\rm{class}}}
\newcommand{\ionein}{i_1,\ldots,i_N}
\newcommand{\pionein}{i_{P1},\ldots,i_{PN}}

\begin{document}



\title{Fully quantum mechanical moment of inertia of a mesoscopic
  ideal Bose gas}

\author{Jens~Schneider\thanks{E-mail: Jens.Schneider@mpq.mpg.de} and
  Hartmut~Wallis\thanks{E-mail: Hartmut.Wallis@t-online.de}}

\address{Max-Planck-Institut f\"ur Quantenoptik,
  Hans-Kopfermann-Stra{\ss}e 1,
  D-85748~Garching and\\
  Sektion Physik, Ludwig-Maximilians-Universit\"at M\"unchen,
  Theresienstra\ss{}e 37, D-80333 M\"unchen, Germany}


\date{preprint, \today}

\maketitle

\begin{abstract} 
  The superfluid fraction of an atomic cloud is defined using the
  cloud's response to a rotation of the external potential, i.e. the
  moment of inertia. A fully quantum mechanical calculation of this
  moment is based on the dispersion of $L_z$ instead of quasi-classical
  averages.
  
  In this paper we derive analytical results for the moment of inertia
  of a small number of non-interacting Bosons using the canonical
  ensemble.  The required symmetrized averages are obtained via a
  representation of the partition function by permutation cycles.  Our
  results are useful to discriminate purely quantum statistical
  effects from interaction effects in studies of superfluidity and
  phase transitions in finite samples.
\end{abstract}
\pacs{03.75.Fi,05.30.Jp}

\section{Introduction}

Despite recent achievements in preparing Bose condensed atomic gases
\cite{ANDE95,BRAD95,DAVIS95} the question of superfluidity still
escapes direct experimental observation.  This is partly due to the
difficulty to define and/or access the appropriate experimental
observables.

Our understanding of superfluidity is formed by the physics of
macroscopic systems such as liquid helium.  In those systems standard
theoretical methods could be applied that required a thermodynamic
limit procedure.

However, for the finite size systems prepared with trap\-ped atomic Bose
gases the standard answers had to be made more precise.  E.g., one can
speak of a phase transition, but a discontinuous change of system
observables does not occur in mesoscopic systems. The phase transition
was linked to the ground state population, but the most striking
effect of superfluidity cannot be observed in a direct way.

In a spatially homogeneous situation the phenomenon of superfluidity
is defined as the suppression of friction for a linear motion slower
than the velocity of sound \cite{e.m.lifshitz80:_statis_physics}. This
phenomenon is constrained to a fraction of the fluid only, the
so-called superfluid fraction. In linear response theory the latter is
calculated via the quantum mechanical dispersion of the momentum
distribution.

The suppression of friction itself can be traced back to the
appearance of excitations with a linear dispersion relation for an
interacting Bose gas in the presence of a condensate. It is thus
inseparably connected with interparticle interactions. Nevertheless
the superfluid fraction mentioned above does not vanish for an ideal
Bose gas due to Bose-Einstein statistics at least for mesoscopic
samples.

In the case of a atomic cloud trapped e.g. in a harmonic potential
a similar qualitative change of the spectrum of low lying excitations
can not be observed, and the understanding of superfluidity in
rotational motion of trapped atoms is more subtle.
Instead of regarding the response to a Galilei shift one has to
consider the response of the gas to rotations. The response
coefficient then is the moment of inertia of the trapped gas.

The approach of Brosens et. al. to the moment of inertia is based on
the classical expectation value $\langle x^2 + y^2\rangle$
\cite{brosens97:_rotat}. Their analysis focuses on the difference in
the moment of inertia of a totally classical Boltzmann gas in a trap
and the expectation value of $\langle x^2+y^2\rangle$ for a Bose gas
(cf.\ Eq.~(\ref{eq:Iclass})). Therefore they miss the true superfluid
effects that may only be analyzed by calculating the moment of inertia
from quantum mechanical response to rotations.

By contrast, Stringari's work \cite{STRI96:2} is based on linear
response theory.  He obtained the different contributions from the
condensate and the thermal cloud to the response coefficient both for
an ideal and an interacting Bose gas. His use of the grand canonical
ensemble is certainly justified for the study of $10^6$ particles, a
characteristic number for present BEC experiments.

Instead of considering the relation between rotations and
superfluidity one might investigate the relation between dissipation
and superfluidity like in
\cite{jackson98:_vortex_bose_einst,raman99:_eviden_bose_einst,%
fedichev00:_critic_bose_einst} where the onset of dissipation is
analyzed depending on the velocity of an external perturbation. The
numerical work in \cite{jackson98:_vortex_bose_einst} favors vortex
creation as the main dissipation mechanism but it is still under
debate at which critical velocity dissipation sets in.

The rotational properties have recently been analyzed in various works
either focusing on vortices
\cite{DALF98:1,matthews99:_vortic_bose_einst,madison00:_vortex_bose_einst}
or on the so-called scissors mode
\cite{guery-odelin99:_sciss_bose_einst,marago00:_obser_bose_einst}.
The analysis of this mode led to a connection between the quadrupole
excitations and the moment of inertia of the normal fluid fraction
\cite{zambelli00:_momen_inert_quadr_respon_funct_trapp_super} which
might open a way to measure the moment of inertia.

In this paper we present a calculation of the fully quantum mechanical
moment of inertia for a mesoscopic cloud of non-interacting atoms in a
cylindrically symmetrical trap. Finite size effects are allowed for by
calculating the canonical ensemble averages appropriate for this
regime. In this respect, our calculations are complementary to
\cite{STRI96:2} and show markedly different results for the superfluid
fraction. It is of particular interest that all relevant averages are
expressed using permutation cycles which have already played a crucial
rule in previous Path-Integral-Monte-Carlo (PIMC) studies
\cite{CEPE95,KRAU96}. Our analytical results are compared to and
corroborated by numerically exact results computed by the PIMC method.

In Section~\ref{sec:canensemb} we first present the method of
permutation cycles and apply it to the evaluation of the moment of
inertia in Section~\ref{sec:supratrap}. In Sec.~\ref{sec:numres}, we
finally compare the results obtained by the different methods.

\section{Canonical averages}
\label{sec:canensemb}


We want to perform our calculations using the permutation cycle
analysis introduced by Feynman \cite{r.p.feynman72:_statis_mechany}
and the canonical ensemble. Let us consider $N$ particles living in a
exterior potential $V(\vecr)$ ($\vecr \in R^D$), so the single
particle Hamiltonian is given as usual by
\begin{equation}
  \label{eq:singleH}
  H = \frac{{\vecp}^2}{2m} + V(\vecr).
\end{equation}
With the help of the eigenvalues $E_i$ and eigenfunctions
$|\phi_i\rangle$ $H$ can also be written as
\begin{equation}
  \label{eq:singleHeigen}
  H = \sum_i E_i |\phi_i\rangle\langle\phi_i|.
\end{equation}
The total Hamiltonian for $N$ particles is then given by the sum $H_N =
\sum_{j=1}^N H^{(j)}$ over all the particles.

The central physical quantity in statistical mechanics is the
partition function $Z_N(\beta)$ at inverse temperature $\beta = 1/(kT)$.
For a gas of $N$ bosons $Z_N(\beta)$ is given by
\begin{equation}
  \label{eq:ZNbasic}
  Z_N(\beta) = \frac{1}{N!} \sum_{P\in S_n} \int \rm{d} R\, \rho(R,PR),
  \qquad R = (\vecr_1,\ldots,\vecr_N)
\end{equation}
where $\rho(R,PR) = \langle R | \E^{-\beta H_N} | PR \rangle$ is the
density matrix between the point $R$ and the permuted point
$PR = (\vecr_{P1},\ldots,\vecr_{PN})$ and $P$ is a permutation of the first $N$
integer numbers. As the total Hamiltonian $H_N$ is a sum of
independent single-particle Hamiltonians the integral factorizes
\begin{equation}
  \label{eq:ZNfactorize}
  Z_N(\beta) = \frac{1}{N!} \sum_{P\in S_n} \int
  \prod_{j=1}^N \rm{d}^D \vecr_j\, \rho_1(\vecr_j, \vecr_{Pj}).
\end{equation}
Here,
$\rho_1(\vecr_, \vecr_{Pj}) = \langle\vecr_j |\E^{-\beta H} | \vecr_{Pj} \rangle$
is the single-particle density matrix. Now, we break up the
permutations into so-called "cycles", that is subsets of the number
from 1 to $N$ that are invariant under the action of a permutation
$P$. If we break up $P$ in this way, we may get $C_q$ cycles of length
$q$; as we are working in the canonical ensemble these numbers are
restricted by $\sum_{q=1}^N q C_q = N$.  Rearranging the integrand of
(\ref{eq:ZNfactorize}) one arrives at
\begin{equation}
  \label{eq:ZNnice}
  Z_N(\beta) = \sumCq
     \prod_q \frac{Z_1(q\beta)^{C_q}}{C_q! q^{C_q}}
\end{equation}
for the partition function (see \cite{r.p.feynman72:_statis_mechany}).
Here, the sum over all combinations of "cycle populations"
${C_1,\ldots,C_N}$ is restricted by $\sum_{q=1}^N q C_q = N$.

By calculating the derivative with respect to $\beta E_i$ one gets the
formula for $\langle N_i \rangle$ for later evaluations (see also
\cite{WEIS97})
\begin{equation}
  \label{eq:ni}
  \langle N_i \rangle = -\frac{1}{Z_N(\beta)}
                         \frac{\partial Z_N(\beta)}{\partial \beta E_i}.
\end{equation}
We now apply this expression to (\ref{eq:ZNnice}) and use the fact
that $Z_1(q\beta) = \sum_i \E^{-q\beta E_i}$ to finally obtain
\begin{equation}
  \label{eq:ninice}
  \langle N_i \rangle = \sum_{q=1}^N
                        \frac{\E^{- q\beta E_i}}{Z_1(q\beta)}
                        \langle q C_q \rangle.
\end{equation}
So we need to know the mean number of q-cycles $\langle C_q \rangle$
to compute $\langle N_i \rangle$.

Evidently, $\langle C_q \rangle$ is defined by
\begin{equation}
  \label{eq:cqprincipa}
  \langle C_q \rangle = \frac{1}{Z_N(\beta)}
  \sumCr \prod_{r=1}^N \frac{Z_1(r\beta)^{C_r}}{C_r! r^{C_r}} C_q.
\end{equation}

To calculate this expression, we split the product into the factors
with $r\neq q$ and the factor with $r=q$. Note also that terms with
$C_q=0$ do not contribute. So one gets
\begin{eqnarray}
  \label{eq:cqsplit}
  \langle C_q \rangle &=&  \frac{1}{Z_N(\beta)}
  \sumCr \nonumber\\
  && \prod_{r=1, r\neq q}^N \frac{Z_1(r\beta)^{C_r}}{C_r! r^{C_r}} 
  \cdot
  \frac{Z_1(q\beta)^{C_q-1}}{(C_q-1)! q^{C_q-1}} \frac{Z_1(q\beta)}{q}.
\end{eqnarray}
As $C_q>0$, we can substitute $C_q-1$ by $C_q$ and do the sum again
from $C_q=0$ to $\infty$. But this means that we consider one
$q$-cycle less, so $\sum_r r C_r = N-q$. We end up with the final
formula for the cycle occupation number
\begin{equation}
  \label{eq:Cqfinal}
  \langle C_q \rangle = \frac{ Z_{N-q}(\beta)}{Z_N(\beta)}
                        \frac{Z_1(q\beta)}{q}
\end{equation}
where $Z_{N-q}(\beta)$ originates from the sum over the products in
(\ref{eq:cqsplit}). This equation together with the constraint on the
$C_q$'s constitute the well-known recursion relations for $Z_N(\beta)$
\cite{WEIS97}.

\section{Suprafluidity in a harmonic trap}
\label{sec:supratrap}

In this section we want to compute the superfluid fraction $\rhost$ of
a gas of noninteracting bosons in a harmonic trap. This will be done
by using the permutation cycles introduced in the last section.

The superfluid fraction can be defined via the response of the system
to infinitesimal rotations just like in the usual case for
translations \cite{e.m.lifshitz80:_statis_physics,CEPE95}. The
superfluid part shows no response to rotations at all while its
density distribution contributes to the classical moment of inertia.
Therefore, one has
\begin{equation}
  \label{eq:rhosnormal}
  \rhos = 1 - \rhon,
\end{equation}
with the normal fluid fraction defined by the quotient of the quantum
mechanical and the classical moment of inertia for rotations around
the symmetry axis ($z$-axis)
\begin{equation}
  \label{eq:rhon}
  \rhon = \frac{\iqm}{\iclass}.
\end{equation}
One can calculate $\iqm$ via the response to rotations \cite{CEPE95},
this yields
\begin{equation}
  \label{eq:Iqm}
  \iqm = \beta
        \lkl \langle L_z^2 \rangle - \langle L_z\rangle^2\rkl
\end{equation}
or
\begin{equation}
  \label{eq:Iqmfinal}
    \iqm = \beta\langle L_z^2 \rangle,
\end{equation}
because we only consider non-rotating situations with
$\langle L_z\rangle = 0$. The classical moment of inertia is defined as usual
by
\begin{equation}
  \label{eq:Iclass}
  \iclass = m \sum_{j=1}^N \langle (x_j^2+y_j^2)\rangle.
\end{equation}

We now want to compute both (\ref{eq:Iqmfinal}) and (\ref{eq:Iclass})
by using the permutation cycles of section \ref{sec:canensemb}.

We consider the single-particle Hamiltonian for a deformed, harmonic
potential in three dimensions
\begin{equation}
  \label{eq:htrap}
  H = \frac{{\vecp}^2}{2 m} + \frac{1}{2}m \lkl \omega_{\bot}^2 (x^2+y^2)
  + \omega_\|^2 z^2 \rkl.
\end{equation}
Its eigenfunctions can be classified by three quantum numbers
$n_r=0,1,2,\ldots, m=0,\pm 1,\ldots, n_z=0,1,\ldots$ with
\begin{eqnarray}
  \label{eq:Heigen}
  H|n_r, m, n_z\rangle &=& \big\{ \hbar\omega_\bot (2n_r + |m| + 1)  \nonumber\\
  & & \qquad + \hbar\omega_\|
  (n_z+1/2) \big\}  |n_r, m, n_z\rangle ;
\end{eqnarray}
they are also eigenfunctions of the angular momentum operator around
the $z$-direction
\begin{equation}
  \label{eq:leigen}
  l_z|n_r, m, n_z\rangle = m\hbar|n_r, m, n_z\rangle.
\end{equation}

The total angular momentum is given by the sum over the angular
momentum operators for the $N$ particles in the trap
\begin{equation}
  \label{eq:Ltotal}
  L_z = \sum_{j=1}^N l_z^{(j)}.
\end{equation}

We first turn to $\langle L_z^2 \rangle$. Instead of expressing the
sum over all states in the thermodynamic averaging as integrals over
the particle positions like in (\ref{eq:ZNbasic}) we here use the
basis of the single-particle states in (\ref{eq:Heigen}) to calculate
the expectation value of $L_z^2$ (we use $i_j=(n_{r,j}, m_j, n_{z,j})$
to denote the states of particle $j$)
\begin{eqnarray}
  \label{eq:lz2.1}
  \langle L_z^2 \rangle &=& \frac{1}{Z_N N!}
      \sum_{P\in S_N}\sum_{\ionein} \nonumber\\
      && \langle{\ionein}|\sum_{j,k=1}^N l^{(j)}_z l^{(k)}_z \E^{-\beta
         H_N}|\pionein \rangle.
\end{eqnarray}
Again, one can factorize the matrix element to a product of matrix
elements for only one particle. The factors are either of the form
$\langle i_j |\E^{-\beta H^{(j)}}| i_{Pj}\rangle = \E^{-\beta E_{i_j}}
\delta_{i_j, i_{P_j}}$ or
$\langle i_j |l^{(j)}_z\E^{-\beta H^{(j)}}|
i_{Pj}\rangle = \hbar m_j \E^{-\beta E_{i_j}} \delta_{i_j, i_{P_j}}$
or
$\langle i_j |{l^{(j)}_z}^2\E^{-\beta H^{(j)}}| i_{Pj}\rangle =
(\hbar m_j)^2 \E^{-\beta E_{i_j}} \delta_{i_j, i_{P_j}}$.

If now the sum of the permutations is expressed as a sum over all
cycle occupations one can see, that due to the Kronecker-$\delta$s in
the factors, all particles on the same $q$-cycle have the same state.
For one $q$-cycle, there are again three different possibilities: if
there is no $l^{(j)}_z$ associated to one of the particles on the
cycle, then (\ref{eq:lz2.1}) gets the contribution
\begin{equation}
  \label{eq:lz2.2}
  \sum_{i_j} \E^{-q\beta E_{i_j}} = Z_1(q\beta).
\end{equation}
Or there may be only one such $l^{(j)}_z$. Then the contribution is
\begin{equation}
  \label{eq:lz2.3}
  \sum_{i_j} \hbar m_j \E^{-q\beta E_{i_j}} = 0,
\end{equation}
which vanishes due to the symmetry $E_{n_r,m,n_z} = E_{n_r,-m,n_z}$.
The third possibility is to have two angular momentum operators acting
on two particles on the same q-cycle. This contributes a factor
\begin{equation}
  \label{eq:lz2.4}
  \sum_{i_j} (\hbar m_j)^2 \E^{-q\beta E_{i_j}} =
  \frac{2\hbar^2 \E^{-q\beta\hbar\omega_\bot}}
                {\lkl 1 - {\E^{-q\beta\hbar\omega_\bot}} \rkl^2}
  Z_1(q\beta).
\end{equation}
So all $q$-cycles contribute a factor of $Z_1(q\beta)$, those
involving two angular momentum operators additionally contribute a
factor of $\frac{2\hbar^2 \E^{-q\beta\hbar\omega_\bot}} {\lkl 1 -
  {\E^{-q\beta\hbar\omega_\bot}} \rkl^2}$.

Before we can write down the formula for $\langle L_z^2 \rangle$ we
must count the number of ways how the two $l_z$-operators may be
distributed among the cycles: for a given number $C_q$ of $q$-cycles
there are $q C_q$ particles sitting on these cycles. They can be
paired with $q$ other particles on their own cycle (including
themselves), so there are $q^2 C_q$ ways to pair the two angular
momentum operators to get (\ref{eq:lz2.4}).  This leads to
\begin{eqnarray}
  \label{eq:lz2prenice}
  \langle L_z^2 \rangle &=& \frac{1}{Z_N N!} \sumCq
  M({C_r}) \sum_{q=1}^N \nonumber\\
  && \frac{2\hbar^2 \E^{-q\beta\hbar\omega_\bot}}
  {\lkl 1 - {\E^{-q\beta\hbar\omega_\bot}} \rkl^2} q^2 C_q
  \prod_{r=1}^N Z_1(r\beta)^{C_r},
\end{eqnarray}
where
\begin{equation}
  \label{eq:MC1CN}
  M(C_1,\ldots,C_N) = \frac{N!}{\prod_q C_q! q^{C_q}}
\end{equation}
is the number of permutations with $C_1$ 1-cycles, $C_2$ 2-cycles etc
(see \cite[above Eq.~(2.154)]{r.p.feynman72:_statis_mechany}).  By
using this and Eqs.~(\ref{eq:cqprincipa}) and (\ref{eq:Iqmfinal})
\begin{equation}
  \label{eq:lz2nice}
  \iqm = \beta \langle L_z^2 \rangle =
  2\hbar^2 \sum_{q=1}^N
  \frac{q\beta \E^{-q\beta\hbar\omega_\bot}}
  {\lkl 1 - {\E^{-q\beta\hbar\omega_\bot}} \rkl^2} \langle q C_q \rangle.
\end{equation}
So we can calculate the quantum mechanical value of the momentum of
inertia by using the cycle occupations in (\ref{eq:Cqfinal}).

Now we turn to the classical moment of inertia. One can do the
analogous analysis as before for $\iqm$. Here, one meets terms like
\begin{equation}
  \label{eq:x2y2}
  \sum_{i_j} \langle i_j|(x_j^2+y_j^2) \E^{-q\beta H}  | i_j \rangle = 
  \frac{\hbar}{m\omega_\bot}
  \frac{1+\E^{-q\beta\hbar\omega_\bot}}
       {1-\E^{-q\beta\hbar\omega_\bot}} Z_1(q\beta)
\end{equation}
in analogy to (\ref{eq:lz2.3}) or (\ref{eq:lz2.4}).
Eq.~(\ref{eq:x2y2}) is most easily computed using the eigenstates
$|n_x,n_y,n_z\rangle$ of the harmonic trap Hamiltonian.

Finally, $\iclass$ can be written as
\begin{equation}
  \label{eq:iclassfinal}
  \iclass = \frac{\hbar}{\omega_\bot} \sum_{q=1}^N
  \frac{1+\E^{-q\beta\hbar\omega_\bot}}
       {1-\E^{-q\beta\hbar\omega_\bot}}
  \langle q C_q \rangle,
\end{equation}
which again depends on the cycle occupation numbers.
Eqs.~(\ref{eq:lz2nice}) and (\ref{eq:iclassfinal}) constitute the main
result of the present work, they allow the computation of the
superfluid fraction from Eqs.~(\ref{eq:rhosnormal}), (\ref{eq:rhon})
totally based on the cycle occupation numbers in (\ref{eq:Cqfinal}).

\section{Numerical results and comparison to path integral Monte-Carlo
  results}
\label{sec:numres}

We now turn to the comparison of the cycle approach with results from
other methods for the description of the trapped Bose gas.

\begin{figure}[htbp]
  \begin{center}
    \epsfxsize=0.47\textwidth
    \epsffile{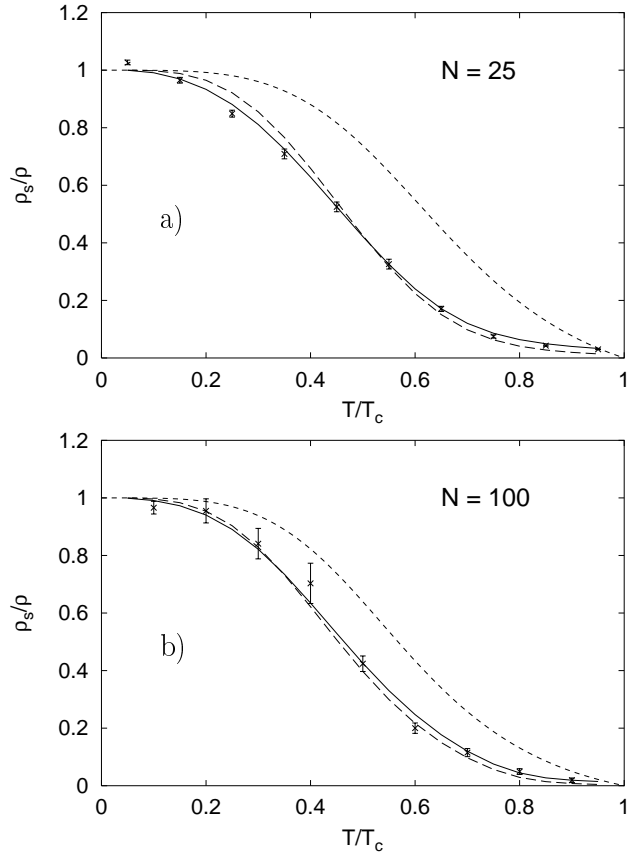}
    \caption{Superfluid density for $N=25$ (a)) resp. $N=100$ (b))
      particles in a spherically symmetric harmonic trap as a function
      of temperature ($T_{\rm c}$ is the usual critical temperature
      for Bose condensation in a harmonic trap). The full line denotes
      our results using the cycle analysis. The short dashed line
      shows the results of Stringari's considerations based on the
      grand canonical ensemble \protect\cite{STRI96:2}.  The long
      dashed curve stems from the modified two-fluid model
      (Eq.~(\ref{eq:rhosest})). The crosses with error bars are from
      a Monte-Carlo calculation based on path integrals.}
    \label{fig:rhosN25100}
  \end{center}
\end{figure}

In Fig.~\ref{fig:rhosN25100} we show the superfluid fraction as a
function of temperature for $N=25$ particles in a spherical trap.  The
full line shows the result originating from our cycle analysis. It is
obtained by using (\ref{eq:rhosnormal}, \ref{eq:rhon},
\ref{eq:lz2nice}, \ref{eq:iclassfinal}). $\langle q C_q \rangle$ has
been calculated from the recursion relations for $Z_N(\beta)$.
The first comparison is with the model of Stringari \cite{STRI96:2}
who has given a formula for $\rhost$ based on grand canonical
considerations. This result is plotted with short dashes in
Fig.~\ref{fig:rhosN25100}.

The difference between the canonical and grand
canonical values is clearly visible both for $N=25$ particles
and for $N=100$. The two chosen examples already illustrate that
the difference between the two ensembles will vanish in the limit $N
\to \infty$.

It is interesting that by using the canonical expectation values we
can also reconcile the cycle analysis given above with a simple
two-fluid model of the inhomogeneous Bose gas
\cite{tisza38,NARA98:1,dodd99:_two_bose_einst}. In that model the
superfluid fraction is totally made up of the condensed part of the
system and the normal fluid is identical to the non-condensed part.
We here modify the two-fluid model by inserting the number of
condensed particles $\langle N_0 \rangle$ (see (\ref{eq:ninice})) from
the canonical averages. For the case of the harmonic trap, $\langle
N_0 \rangle$ can be written as
\begin{equation}
  \label{eq:nnulltrap}
  \langle N_0 \rangle = \sum_{q=1}^N
             \lkl 1-\E^{-q\beta\hbar\omega_\bot}\rkl^2
             \lkl 1-\E^{-q\beta\hbar\omega_\|}\rkl
             \langle q C_q \rangle.
\end{equation}
To compute $\rhost$ we need the moments of inertia $I_0$ of the
condensate and $I_{\rm nc}$ of the non-condensed part. The
condensate particles all reside in the ground state of the trap whose
moment of inertia for rotations around the $z$-axis is
$\hbar/\omega_\bot$, so
\begin{equation}
  \label{eq:i0}
  I_0 = \langle N_0 \rangle \frac{\hbar}{\omega_\bot}.
\end{equation}
$I_{\rm nc}$ is estimated by assuming that the non-condensed
particles behave like a Boltzmann gas in the harmonic trap. For the
inverse temperature $\beta$ the moment of inertia then equals
$2/(\beta\omega_\bot^2)$ so
\begin{equation}
  \label{eq:inc}
  I_{\rm nc} = N_{\rm nc} \frac{2}{\beta\omega_\bot^2}
  = N_{\rm nc} \frac{2kT}{\omega_\bot^2},
\end{equation}
where $N_{\rm nc} = N-\langle N_0 \rangle $.

Finally, by noting that the condensate only contributes to the
classical moment of inertia $\iclass$ but does not take part in the
rotation, we can estimate $\rhost$ by
\begin{eqnarray}
  \label{eq:rhosest}
  \rhos &\approx& 1 - \frac{I_{\rm nc}}{I_0 + I_{\rm nc}}
    = \frac{1}{1 + \frac{N-\langle N_0 \rangle}{\langle N_0 \rangle}
      \frac{2 kT} {\hbar\omega_\bot}}.
\end{eqnarray}
The long dashed lines in Fig.~\ref{fig:rhosN25100} give a plot of this
formula. It fits the exact result for the canonical ensemble
surprisingly well. The two-fluid model differs from the exact result
only due to the small difference between the true non-condensed part and
its quasi-classical approximation. 

A third way to calculate the superfluid fraction via moments
of inertia is the path integral representation of the density matrix
\cite{r.p.feynman72:_statis_mechany}. This approach works for both
non-interacting and interacting systems. It represents the most
important point of comparison as it is a potentially exact method. We
have implemented a Path Integral Monte-Carlo (PIMC) code that relies
on the bisectioning ideas of Ceperley \cite{CEPE95} and on the
factorization of the complete density matrix into a non-interacting
part and the interaction correction \cite{KRAU96}.

Here, we will only show that the results discussed in the previous
section agree well --- as they should --- with the data obtained from
PIMC calculations without interactions.

The data points (crosses) in Fig.~\ref{fig:rhosN25100} show our
results for the superfluid fraction. They have been computed by using
the so-called ``area formula'' \cite{SIND89,CEPE95} which is the most
appropriate method for our investigations. The superfluid fraction is
obtained from
\begin{equation}
  \label{eq:pimcrhos}
  \frac{\rho_s}{\rho} = \frac{4 m^2}{\hbar^2\beta}
  \frac{\langle A_z^2 \rangle}{\langle I_z\rangle}.
\end{equation}
Here
\begin{equation}
  \label{eq:pimcIz}
  I_z = \frac{m}{M}\sum_{t=0,i=1}^{M-1,N}
        \big( x_i(t)x_i(t+1) + y_i(t)y_i(t+1)\big)
\end{equation}
denotes the PIMC approximation of classical moment of inertia.  $m$ is
again the mass of the particles. $x_i(t)$ and $y_i(t)$ are the
coordinates of the $i$-th particle on time slice $t$ of the PIMC
simulation and there are $M$ such time slices.  $I_z$ clearly
converges to $\iclass$ as $M\to\infty$.

The expression
\begin{equation}
\label{eq:pimcAz}
   A_z = \frac{1}{2}\sum_{t=0,i=1}^{M-1,N}
          \big( x_i(t)y_i(t+1) - y_i(t)x_i(t+1)\big)
\end{equation}
is the projected area perpendicular to the rotation axis $z$.
$\langle A_z^2 \rangle$ is the portion of the moment of inertia that
can be traced back to the superfluid fraction \cite{CEPE95}.

As Fig.~\ref{fig:rhosN25100} shows, all three methods are in good
agreement with each other. We have furthermore calculated the density
distribution of the particles in the trap both in the two-fluid model
and with PIMC. They also exhibit a nice agreement thus indicating the
validity of the empirical two-fluid model.

\section{Summary and Conclusion}
\label{sec:summ}

The main result of our paper is the calculation of the superfluid
fraction from the permutation cycles. We have compared this approach
to the grand-canonical prediction by Stringari and to PIMC
calculations. For small particle numbers our results are in good
agreement with the exact (canonical) PIMC results and we were able to
reproduce them to a very good accuracy with a two-fluid model which
divides the gas into a condensed and a non-condensed part where the
latter is treated as a classical Boltzmann gas. For small particle
numbers we find a distinct difference between our results and
Stringari's grand canonical approach.

The techniques and results presented in this paper have established a
solid starting point of PIMC investigations including interactions.
As the calculation of the condensate fraction in PIMC calculations of
inhomogeneous Bose gases is still under debate, the role of the
permutation cycles deserves further investigations also in the
interacting case (see e.g. \cite{NORD97} for a related discussion).

\acknowledgements
J.S. thanks M.~Holzmann for a stimulating discussion on the subject.
We gratefully acknowledge financial support by DFG under Grant Nr.
SCHE~128/7-1.


\end{document}